\documentclass[twocolumn,9pt]{article} 

\usepackage[square,numbers,sort&compress,comma]{natbib}

\usepackage[square,numbers,sort&compress,comma]{natbib}
\usepackage{amssymb}
\usepackage{amsmath}
\usepackage{xcolor}
\usepackage{amsmath}
\usepackage{amssymb}
\usepackage{caption}
\usepackage{graphicx}
\usepackage{latexsym}
\usepackage{times}
\usepackage[utf8]{inputenc}
\usepackage[hyphens]{url}
\usepackage{hyperref}
\usepackage{multirow}
\usepackage{booktabs}
\usepackage[version=4]{mhchem}
\usepackage{siunitx}
\DeclareSIUnit\pixel{px}
\topmargin - 12pt 
\renewenvironment{abstract}%
              {
               \small
               {\bfseries \abstractname}
               \par
               \vspace{10pt}
              }

\renewcommand\abstractname{Abstract}

\newcommand{\nomenclature}
              [1]
              {
               \bgroup
               \flushleft
               \small\bf
               #1
               \par
               \egroup
              }

\renewcommand{\section}
              [1]
              {
               \bgroup
               \flushleft
               \small\bf
               \refstepcounter{section}
               \arabic{section}. #1
               \par
               \egroup
              }

\renewcommand{\subsection}
              [1]
              {
               \bgroup
               \flushleft
               \small\em
               \refstepcounter{subsection}
               \arabic{section}.
               \arabic{subsection}. #1
               \par
               \egroup
              }

\renewcommand{\subsubsection}
              [1]
              {
               \bgroup
               \flushleft
               \small\em
               \refstepcounter{subsubsection}
               \arabic{section}.
               \arabic{subsection}.
               \arabic{subsubsection}. #1
               \par
               \egroup
              }

  \newcommand{\acknowledgement}
              [1]
              {
               \bgroup
               \flushleft
               \small\bf
               #1
               \par
               \egroup
              }

  \newcommand{\sectionbib}
              [1]
              {
               \bgroup
               \flushleft
               \small\bf
               #1
               \par
               \egroup
              }

\begin{document}



\small
\baselineskip 10pt

\setcounter{page}{1}
\title{\LARGE \bf  Unified scaling and shape laws for turbulent premixed methane
and hydrogen jet flames} 

\author{\large Aurora Maffei$^{a,\dagger}$, Thomas L.~Howarth$^{b,a,\dagger,*}$, Marianna Cafiero$^{a}$, Florence Cameron$^{a}$\\Michael Gauding$^{a}$, Joachim Beeckmann$^{a}$, Heinz Pitsch$^{a}$\\[10pt]
{\footnotesize \em $^a$RWTH Aachen University, Institute for Combustion Technology, Templergraben 64, 52056 Aachen, Germany}\\[-5pt]
{\footnotesize \em $^b$Department of Aeronautical and Automotive Engineering, Loughborough University, Loughborough LE11 3TU, UK}\\[-5pt]
}

\date{}  

\twocolumn[\begin{@twocolumnfalse}
\maketitle
\rule{\textwidth}{0.5pt}
\vspace{-5pt}

\begin{abstract} 

The scaling of turbulent premixed flames is typically described by correlations derived for unity-Lewis-number fuels. However, their validity for hydrogen (\ce{H2}) remains uncertain due to the thermodiffusive effects associated with its low Lewis number. 
In this study, turbulent premixed \ce{H2} and methane (\ce{CH4}) jet flames are systematically compared over a wide range of operating conditions. Experiments were conducted for Reynolds numbers between 5,000 and 60,000 and effective Karlovitz numbers spanning 3–368. Flame structure and global flame geometry were characterized using spatially resolved \ce{OH}\textsuperscript{*} chemiluminescence imaging, allowing consistent comparison between the two fuels across different turbulence intensities. The results are interpreted via a unified framework that incorporates two thermodynamic- and fuel-dependent parameters: a flame speed factor, $\alpha$, representing the enhancement of local burning rates, and a shape factor, $\gamma$, describing the scaling of mean flame geometry. 
Despite significant fuel-specific thermodiffusive effects associated with preferential diffusion and intrinsic reactivity, which lead \ce{H2} flames to exhibit enhanced sensitivity to turbulence and more compact flame configurations, both \ce{H2} and \ce{CH4} flames are found to exhibit robust and consistent turbulent scaling behavior when analysed within the proposed unified framework. The resulting correlations provide a generalised description of turbulent burning velocity and flame structure, demonstrating that key turbulence–chemistry interactions can be captured within a common model across fuels with widely different Lewis numbers. Overall, the dataset spans multiple turbulence regimes and flame geometries for both fuels, providing a valuable experimental benchmark for the validation of turbulent combustion models across different regimes.

\end{abstract}

\vspace{10pt}

{\bf Novelty and significance statement}

\vspace{10pt}

This study presents a systematic experimental comparison of turbulent premixed hydrogen and methane jet flames across a wide Reynolds–Karlovitz range. The key novelty is a unified scaling framework governed by two constant fuel-dependent prefactors: a flame-speed factor, $\alpha$, that scales the turbulent burning rate, and a shape factor, $\gamma$, that scales the mean flame geometry. By defining these as invariant prefactors, we demonstrate that both fuels obey the same underlying dynamic scaling laws once fuel-specific constants are applied. These results provide correlations that account for differential diffusion effects, enabling the validation of RANS/LES models without case-specific tuning. Finally, the work establishes the limits of regime validity and provides benchmark data to develop turbulence-flame interaction closures for low-carbon fuels.

\vspace{5pt}
\parbox{1.0\textwidth}{\footnotesize {\em Keywords:} Hydrogen combustion; Premixed flames; Turbulent combustion; Chemiluminescence.}
\rule{\textwidth}{0.5pt}
*Corresponding author. $^{\dagger}$These authors contributed equally.\vspace{5pt}
\end{@twocolumnfalse}] 

\section{Introduction\label{sec:introduction}} 

Hydrogen (\ce{H2}) is increasingly regarded as a key energy carrier for carbon-free combustion technologies. Its use as a primary or blended fuel offers the potential to significantly reduce greenhouse gas emissions while preserving high specific energy content and fast chemical kinetics. However, compared to conventional hydrocarbon fuels, the distinctive properties of \ce{H2}, such as high diffusivity, low ignition energy, and wide flammability limits, introduce significant operational challenges and additional complexity in control and modeling~\cite{bouvet2013effective, pitsch2024}. In particular, the low Lewis number ($Le$) of \ce{H2} enhances differential and preferential diffusion, which can trigger thermodiffusive instabilities (TDI) leading to cellular flame structures and local variations in burning rate~\cite{haq2002wrinkling, bell2007numerical,day2009turbulence,aspden2011turbulence,howarth2023thermodiffusively}. 
The interplay between turbulence, diffusion, and chemical kinetics complicates the application of classical turbulent combustion scaling laws that were developed for $Le\approx1$ fuels, such as methane (\ce{CH4})~\cite{Plessing2000,Vargas2020}. 
In turbulent premixed combustion, these interactions are typically characterized through non-dimensional parameters, such as the Reynolds ($Re$), Damk\"ohler ($Da$), and Karlovitz ($Ka$) numbers~\cite{damkohler1940einfluss, peters1999turbulent}, which describe regimes from flamelet to thin-reaction-zone and distributed combustion.

For fuels with $Le \approx 1$, numerous studies have established correlations linking global quantities such as turbulent burning velocity, flame length, and surface area to key dimensionless numbers~\cite{driscoll2008turbulent, gulder1991turbulent}. In contrast, for lean premixed \ce{H2}–air flames with $Le<1$, both experiments and simulations have highlighted the strong influence of differential diffusion and TDI on local burning rates and flame structure~\cite{ASPDEN20111463, ASPDEN20171997, dinkelacker2011modelling, troiani2024scaling}. Despite extensive investigations into lean premixed \ce{H2}/air jet flames over the years~\cite{wu1990turbulent,chaib2024experimental,shi2024internal}, most studies have focused on single fuel compositions or fixed configurations. Consequently, it remains unclear how local instabilities manifest macroscopically, leaving the global scaling and mean shape of turbulent premixed \ce{H2} flames less understood and the validity of unity-Lewis-number correlations under thermodiffusively unstable conditions uncertain.

The present work addresses this gap through a systematic experimental comparison of turbulent \ce{CH4}/air and \ce{H2}/air premixed jet flames across a wide range of $Re$, $Ka$, and $Da$. By utilizing a unity-Lewis-number reference fuel under identical conditions, this study evaluates whether TDI-induced modifications follow a systematic trend or represent a fundamental breakdown of classical scaling laws. Global flame characteristics, extracted from OH\textsuperscript{*} chemiluminescence imaging, are interpreted through a unified framework introducing two empirical parameters: the flame speed factor $\alpha$, quantifying burning rate modifications between the two fuels, and the shape factor $\gamma$, describing mean flame geometry.

The paper is structured as follows: Section~\ref{subsec:theory} presents the theoretical background underlying turbulent premixed flame scaling and the derivation of the proposed $\alpha-\gamma$ framework, followed by a description of the experimental setup and diagnostic methodology in Section~\ref{subsec:expsetup}. Section~\ref{subsec:st} discusses the results in terms of turbulent burning velocity and determining the values of $\alpha$. Section~\ref{subsec:hf_s} analyzes the mean flame geometry, providing a power law description of the mean flame shape with geometric coefficient $\gamma$. This shape model is combined with the turbulent burning velocity model to provide a flame length model. Finally, Section~\ref{sec:end} provides a discussion and overall conclusion of the results, as well as future research directions.
\\
\\

\section{Methodology\label{sec:met}}

\subsection{Theoretical background\label{subsec:theory}} 

Turbulent premixed flame propagation can be interpreted as the interaction between flame-surface wrinkling and local burning velocity. 
Damk\"ohler~\cite{damkohler1940einfluss} proposed two limiting cases of turbulent burning velocities for turbulent premixed flames, namely the small- and large-scale limits. In the large-scale limit (LSL), surface area enhancement is purely kinematic, giving the scaling law
\begin{equation}
    \frac{s_{T}}{s_{L}} \sim \frac{u^{\prime}}{s_{L}},
\end{equation}
where $s_{T}$ is the turbulent burning velocity, $u^{\prime}$ is the root-mean-square (r.m.s.) of the velocity fluctuation, and $s_{L}$ is the unstretched laminar burning velocity. In the small-scale limit (SSL), Damk\"ohler argued that a turbulent diffusion can be substituted for molecular diffusion, leading to the scaling law
\begin{equation}
   \frac{s_{T}}{s_{L}} \sim\sqrt{\frac{u^{\prime}\ell}{s_{L}l_{F}}} = Re_{F}^{-\frac{1}{2}}Re_{t}^{\frac{1}{2}},
\end{equation}
where $\ell$ is the integral length scale, $l_{F}$ is the unstretched flame thickness, $Re_{F} = s_{L}l_{F}/\nu$ is the flame Reynolds number (with kinematic viscosity $\nu$), and $Re_{t} = u^{\prime}\ell/\nu$ is the turbulent Reynolds number. Using a level-set approach, Peters \cite{peters1999turbulent} derived a turbulent burning velocity correlation that covers the entire range and satisfied Damk\"ohler's large- and small-scale limits, given by
\begin{equation}
    \frac{s_{T}}{s_{L}} = 1 - \frac{a_{4}b_{3}^{2}}{2b_{1}}\frac{\ell}{l_{F}}+\sqrt{\left(\frac{a_{4}b_{3}^{2}}{2b_{1}}\frac{\ell}{l_{F}}\right)^{2} + a_{4}b_{3}^{2}\frac{u^{\prime}\ell}{s_{L}l_{F}}},
\end{equation}
where $s_{T}$ is the global turbulent burning velocity, and $a_{i},b_{i}$ are model constants. Importantly, in the limits of small and large Damk\"ohler numbers ($Da = s_{L}\ell /(u^{\prime}l_{F})$), two distinct scaling behaviors emerge
\begin{equation}
    \frac{s_{T}-s_{L}}{u^{\prime}} = \begin{cases} C_{\rm{SSL}}Da^{\frac{1}{2}}, \quad \text{for}\, Da \ll Da_{0}\\ C_{\rm{LSL}}, \quad \text{for}\, Da \gg Da_{0}\end{cases},
\end{equation}
where $Da_{0}$ denotes the transition Damk\"ohler number separating the small- and large-scale limits and is typically of order unity. Here, the characteristic turbulent velocity fluctuation and integral length scale are assumed to scale with the jet bulk velocity $u_j$ and nozzle diameter $d_j$, respectively. The corresponding flame properties are defined using the rescaled flame speed of a laminar freely propagating flame $s_L^*$ and characteristic flame thickness $l_F^*$. For methane, these are taken as the unstretched 1D quantities ($s_L, l_F$), whereas for hydrogen flames, $s_L^*$ and $l_F^*$ are characterized using the model proposed by \cite{howarth2023thermodiffusively} to account for thermodiffusive effects from the laminar flame. Such a normalization has been deemed necessary by existing work \cite{ASPDEN20111463,aspden2011turbulence}. Introducing these into this model gives the correlations
\begin{equation}\label{eq:jet_st_model}
    \frac{s_{T}}{u_{j}} = \begin{cases} \frac{s_{L}^{*}}{u_{j}} + \alpha_{\rm{SSL}} Re_{F}^{\frac{1}{2}}Re_{j}^{-\frac{1}{2}}\frac{d_{j}}{l_{F}^{*}}, \; \mathrm{SSL} \\ \frac{s_{L}^{*}}{u_{j}} + \alpha_{\rm{LSL}}, \; \mathrm{LSL}\end{cases} ,
\end{equation}
where $Re_{j}$ is the jet Reynolds number ($Re_{j} = u_{j}d_{j}/\nu$) and $\alpha$, introduced here as a flame speed parameter, is given by
\begin{equation}\label{eq:Re_rel}
\alpha_{\rm{S/LSL}} = C_{\rm{S/LSL}}\left(\frac{Re_{t}}{Re_{j}}\right)^{\frac{1}{2}} . 
\end{equation}
Note that $Re_{F}$ is largely independent of whether $^{*}$ quantities have been used, as the product of speed and length scales remains essentially constant. 
The examination of these correlations, in the small-scale limit, is presented in Section \ref{subsec:st}. 

In a turbulent premixed jet flame, the turbulent flame speed can be computed as~\cite{driscoll2008turbulent}
\begin{equation}\label{eq:st_exp}
    s_{T} = \frac{\dot{m}}{\rho_{u}A_{0}} = \frac{u_{j}A_{j}}{A_{0}} , 
\end{equation}
where $\dot{m}$ is the mass flow rate at the inlet, $A_{j}$ is the area of the nozzle and $A_{0}$ is the  cross-sectional area of the mean flame surface, whose evaluation is described in Section \ref{subsec:expsetup}. A relation can be drawn between the normalized flame length, $h_{f}/d_{j}$, and the (inverse) normalized burning velocity, $u_{j}/s_{T}$, as
\begin{equation}\label{eq:st_hf}
    \frac{h_{f}}{d_{j}} = \beta\frac{A_{0}}{A_{j}} = \beta\frac{u_{j}}{s_{T}} , 
  \end{equation}
where $\beta$ is a geometric coefficient that depends on the mean shape of the flame. If the flame shape is fitted with a power law according to
\begin{equation}
    y = h_{f} \left( 1-\left|\frac{x}{r_{j}}\right|^{p} \right) ,
\end{equation}
where $r_{j}$ is the radius of the jet, then, provided that the flame is sufficiently tall, the ratio is given by
\begin{equation}
    \frac{h_{f}}{d_{j}} = \frac{1+p}{4p}\frac{A_{0}}{A_{j}} . 
\end{equation}
Note that particular power laws exhibit different ratios. For example, a cone ($p = 1$) gives a ratio of $1/2$, a parabola ($p=2$) gives $3/8$, and as the flame transitions to a top-hat profile (i.e.~$p\rightarrow\infty$) this limit tends to $1/4$. In reality, due to the slight bulging of the flame, this prefactor is unlikely to exactly match the prescribed function given here. However, this serves as a useful analysis tool for explaining and modeling the relationship between flame shape, turbulent flame speed, and flame length. An examination of this shape factor is performed in Section \ref{subsec:hf_s}.

\subsection{Experimental methodology\label{subsec:expsetup}} 

\begin{figure}[!htbp]
\centering
\includegraphics[width=\columnwidth]{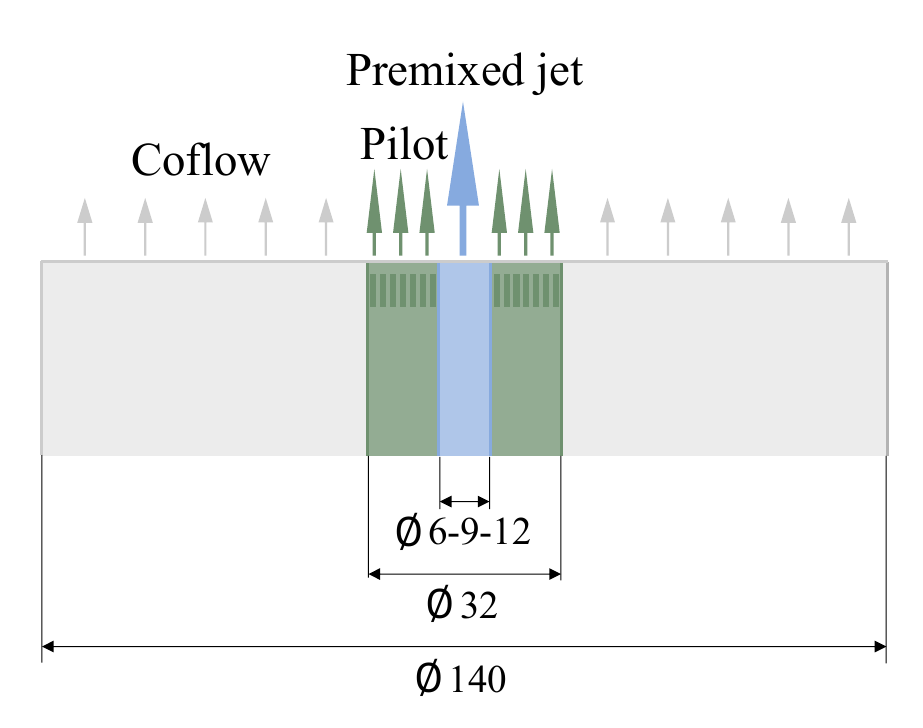}
\caption{Schematics of the round turbulent jet burner showing the central premixed jet surrounded by a pilot and coflow. }
\label{fig:Jet_Burner_fig}
\end{figure}

The experimental setup, shown in Fig.~\ref{fig:Jet_Burner_fig}, consists of a round jet burner operated in an open laboratory environment. The system comprises a central fully premixed fuel-air jet, stabilized by a surrounding laminar pilot flame and enclosed by an outer air coflow shielding. The coflow ring has an inner diameter of \qty{140}{\milli\meter}, while the pilot premixed flame is supplied through a concentric nozzle with an inner diameter of \qty{32}{\milli\meter} and stabilized on a \qty{7}{\milli\meter}-thick alumina honeycomb positioned \qty{3}{\milli\meter} below the main jet nozzle. 
Within the pilot nozzle, two additional concentric nozzles deliver the main reactants: a central fuel nozzle with an inner diameter of \qty{2.5}{\milli\meter} and an annular oxidizer nozzle with interchangeable inner diameters ($d_{j}$) of \qtylist{6; 9;12}{\milli\meter}, depending on the test condition. The fuel nozzle features an adjustable length, enabling variation of the premixing distance and, consequently, control over the degree of premixing. In this study, the burner was operated in fully premixed mode, with an upstream mixing length of approximately \qty{350}{\milli\meter} to ensure complete fuel–oxidizer mixing.

OH\textsuperscript{*} chemiluminescence imaging was performed using an Andor iStar\textsuperscript{\textregistered} ICCD camera coupled with a UV lens and an interferential bandpass filter (310 ± 10 nm). The camera was configured with a gain of 3200, a gate width of 0.005 s, and an exposure time of 0.01 s. Using a multi-shot technique, 700 frames per case were captured at 5 Hz. A background image, acquired daily, was subtracted to remove noise contributions. A convergence analysis on the mean chemiluminescence field confirmed that averaging 700 images was sufficient to achieve a statistically converged flame representation, minimizing the effect of cycle-to-cycle fluctuations. The spatial calibration provided a resolution of \qty{14.75}{\micro\meter\per\pixel}. 


A three-class multi-Otsu thresholding method~\cite{otsu1979} was applied to segment each OH\textsuperscript{*} chemiluminescence intensity field. Before segmentation, all images were normalized to their maximum intensity to minimize the influence of shot-to-shot variations. The method selected two optimal threshold levels that maximize the inter-class variance, providing an adaptive criterion suitable for varying signal intensities. The higher threshold value was used to identify the flame boundary, from which the flame length ($h_{f}$) was determined as the axial distance between the exit of the nozzle surface and the maximum threshold of the OH\textsuperscript{*} emission. The flame surface area ($A_{0}$), on the other hand, was calculated as the area enclosed by the corresponding isoline derived from the segmentation method, after excluding the contribution of the pilot flame. This procedure was applied to all the 700 recorded frames for each operating condition, and $h_{f}$ and $A_{0}$ were obtained by temporal averaging. This approach allowed for a more consistent quantification of both $h_{f}$ and $A_{0}$, especially with better quality in the low-intensity region, and ensured a robust characterization of the overall structure of the flame. It is worth noting that the flame surface area estimated from OH\textsuperscript{*} chemiluminescence represents a line-of-sight integrated projection of the three-dimensional emission, leading to a slight overestimation of the true surface compared to PLIF measurements~\cite{ma2016comparison}. Nevertheless, the scaling relationships discussed in Section~\ref{sec:results} demonstrate that this overestimation is systematic and consistent across all tested conditions. Since the present analysis focuses on relative scaling trends rather than absolute surface values, the bias introduced by line-of-sight integration does not affect the extracted scaling exponents or the fuel-to-fuel comparison. The standard deviation of the measured $h_{f}$ and $A_{0}$, expressed as the standard deviation ($\sigma_{st}$) over 700 frames per operating point, ranges from approximately 4\% for the jet nozzle, $d_{j}$, 12 mm to 2\% for the $d_{j}=6$ mm for $h_{f}$, and from 3\% ($d_{j}= 12 $ mm) to 1.5\% ($d_{j}= 6$ mm) for $A_{0}$. The lower $\sigma_{st}$ observed for the 6 mm nozzle results from the higher jet velocity required to maintain a constant $Re$. The higher velocity results in increased temporal averaging of the flame fluctuations due to the fixed camera exposure time ($t_{\text{exp}} = \qty{0.01}{\second}$). Consequently, the observed reduction in frame-to-frame variability represents as a diagnostic artifact, effectively acting as a low-pass filter, rather than an intrinsic decrease in flame unsteadiness. This effect is observed for both \ce{CH4} and \ce{H2} flames. 
The experimental conditions are shown in Table~\ref{tab:conditions}. A series of premixed \ce{CH4}/air and \ce{H2}/air jet flames was investigated. The main flow consisted of a fully premixed reactant mixture surrounded by an annular pilot flame of identical composition. The global equivalence ratio was set at $\phi = 1$ for \ce{CH4} and $\phi = 0.4$ for \ce{H2} flames. The Reynolds number of the premixed stream, $Re_{j}$, was varied from 5000 to 35000 for \ce{CH4} and from 5000 to 60000 for \ce{H2}, using increments of 5000, with an additional intermediate case at $Re_{j}=7500$ for both fuels. For \ce{CH4}, the exploration of higher $Re_{j}$ values was limited by the onset of blow-off. The pilot stream was supplied at a constant velocity of \qty{0.5}{\meter\per\second} for \ce{CH4}, and \qty{2.0}{\meter\per\second} for \ce{H2}, with their composition and equivalence ratio kept constant throughout the experimental campaign. A low-velocity air coflow (0.2 m/s) was introduced coaxially around the pilot flame to isolate the reacting region from ambient disturbances and minimize external air entrainment. All experiments were conducted at atmospheric pressure (P = \qty{101.3}{\kilo\pascal}) and room temperature ($T\approx\qty{298}{\kelvin}$). The pilot and coflow operating parameters were kept constant in all tests, ensuring reproducibility and isolating the effects of $Re_{j}$ and $d_{j}$. An effective Karlovitz number is calculated as 
\begin{align}
    Ka_{\mathrm{eff}}^{*} &= I_{t}Ka_{j}^{*}, \\
    Ka_{j}^{*} &= \left( u_{j}/s_{L}^{*} \right)^{3/2}\left( d_{j}/l_{F}^{*} \right)^{-1/2},
\end{align}
where $s_{L}^{*}$ and $l_{F}^{*}$ are taken from laminar reference flames at the corresponding equivalence ratio $\phi_{j}$ and considering the model from~\cite{howarth2023thermodiffusively}. $I_{t} = 0.1$ reflects $u^{\prime} \approx 0.1 u_{j}, \ell \approx 0.1 d_{j}$; this constant is unlikely to provide an exact match, but provides a useful order of magnitude.

\begin{table}[h!] 
\centering
\caption{Experimental operating conditions.}
\label{tab:conditions}
\resizebox{\columnwidth}{!}{
\begin{tabular}{lcccccc}
\hline
\toprule
Fuel & $d_{j}$ & $\phi_{j}$ & $Re_{j}$ & $u_{j}$ & $Ka_{j}^{*}$ & $Ka_{\mathrm{eff}}^{*}$\\
--   & [\unit{\milli\meter}]          & --              & $\times10^{-3}$           & [\unit{\meter\per\second}]        & -- & --        \\
\midrule
\hline
\multirow{3}{*}{\ce{CH4}}
 & 6  & \multirow{3}{*}{1.00} & \multirow{3}{*}{$5{-}35$}  
 & 12.7--88.6 & 140-940 & 14--94 \\
 & 9  &     &     & 8.4--59.1 & 60--420 & 6--42 \\
 & 12 &     &     & 6.3--44.3  & 30-240 & 3--24 \\
\addlinespace
\multirow{3}{*}{\ce{H2}} & 6  & \multirow{3}{*}{0.40} & \multirow{3}{*}{$5{-}60$} & 13.5--161.5 & 200--3680 & 20--368 \\
 & 9  &     &     & 9.0--107.7 & 100--1630 & 10--163 \\
 & 12 &     &     & 6.7--80.8  &  50--920 & 5--92 \\
\hline
\bottomrule
\end{tabular}
}
\end{table}

\section{Results and discussion\label{sec:results}} 

\begin{figure}[h!]
\centering
\includegraphics[width=1\columnwidth]{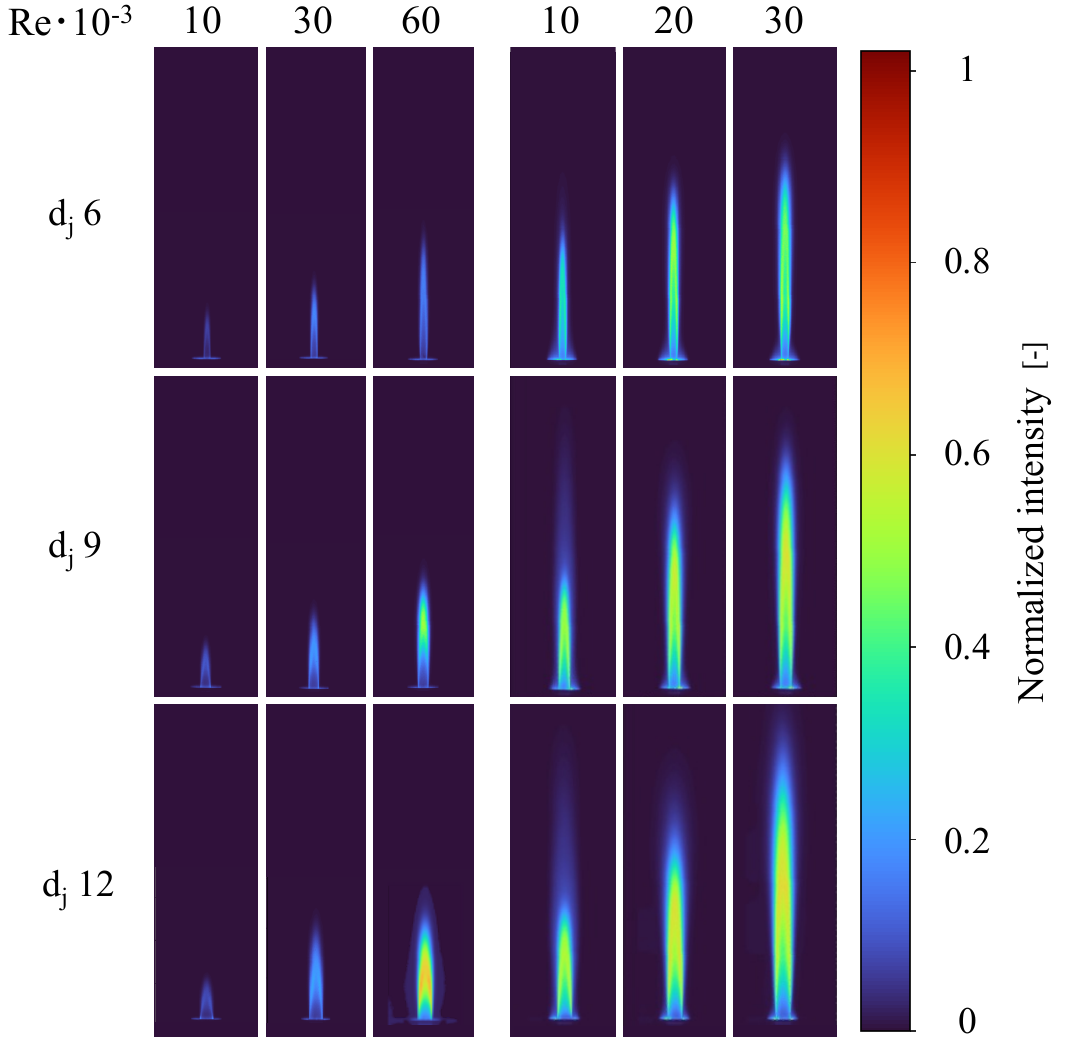}
\caption{ Ensemble-averaged OH\textsuperscript{*} chemiluminescence showing the effect of varying $Re_{j}$ number and injector diameter for \ce{H2} (left) and \ce{CH4} (right). Each fuel was tested at different $Re_{j}$ levels and three diameters ($d_{j}$ = 6, 9, 12\unit{\milli\meter}).}
\label{fig:oh_images}
\end{figure}

The ensemble-averaged OH\textsuperscript{*} chemiluminescence fields over the investigated operating range are shown in Fig.~\ref{fig:oh_images}. For both fuels, increasing the jet Reynolds number, $Re_j$, drives a systematic transition from conical to columnar
flame structures, with this morphological change being significantly more pronounced for \ce{CH4}. The following sections analyze the flame characteristics through the turbulent flame speed and flame shape, which are subsequently combined to model the flame length. 

\subsection{Turbulent burning velocity\label{subsec:st}}

\begin{figure} [h!]
    \centering
    \includegraphics[width=0.85\linewidth]{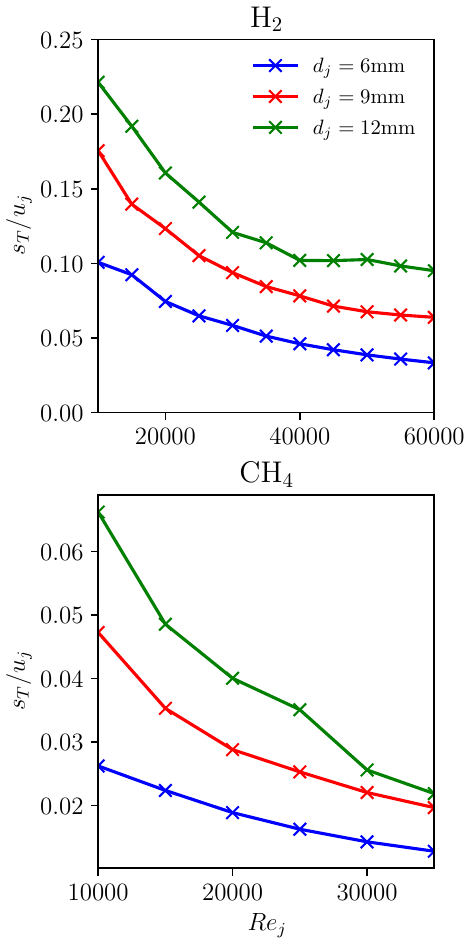}
    \caption{Normalized turbulent burning velocity as a function of $Re_{j}$ for \ce{H2} (top) and \ce{CH4} (bottom) flames.}
    \label{fig:sT}
\end{figure}

Fig.~\ref{fig:sT} presents the normalized turbulent burning velocity, ${s_{T}}/{u_{j}}$, evaluated from the experimental data via Eq.~\eqref{eq:st_exp}, for each of the three diameters as a function of $Re_{j}$ for the \ce{H2} and \ce{CH4} flames. 

\begin{figure} [h!]
    \centering
    \includegraphics[width=0.85\linewidth]{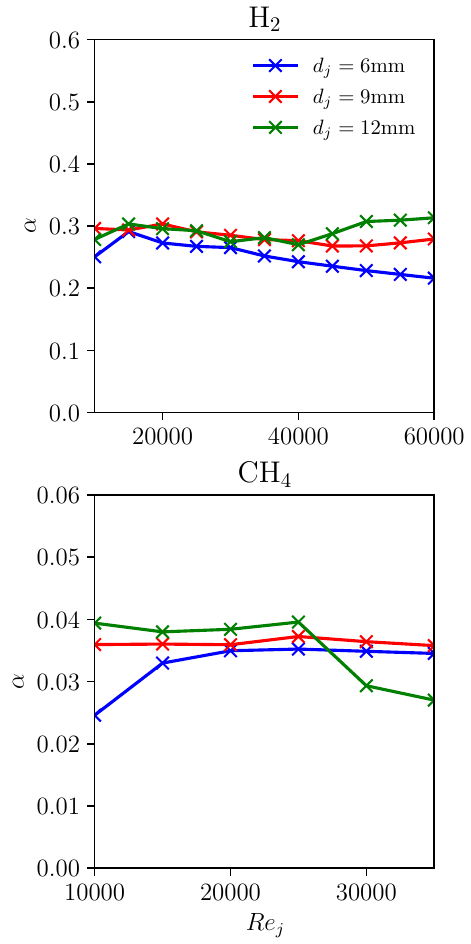}
    \caption{Scaling parameter $\alpha_i$ as a function of $Re_j$ for \ce{H2} (top) and \ce{CH4} (bottom).}
    \label{fig:alpha}
\end{figure}

Each case appears to still be Reynolds-number-dependent, which is unsurprising since the largest effective Damk\"ohler number is around 2 in the $\ce{CH4}, Re_j = 5000, d_{j} = \qty{12}{\milli\meter}$ case, indicating that all of the flames are in the small-scale regime. Under this assumption, Eq.~\eqref{eq:jet_st_model} can be rearranged for $\alpha$

\begin{equation}
    \alpha = \frac{s_{T}-s_{L}^{*}}{u_{j}}Re_{F}^{-\frac{1}{2}}Re_{j}^{\frac{1}{2}}\frac{l_{F}^{*}}{d_{j}}.
\end{equation}
The speed factors for all cases are shown in Fig.~\ref{fig:alpha}. For each fuel, data across the full range of $Re_{j}$ and $d_{j}$ considered here are represented well using a single fuel-specific value of $\alpha$, with $\alpha_{\ce{H2}} \approx 0.28$ and $\alpha_{\ce{CH4}} \approx 0.036$. This order-of-magnitude difference in model constant indicates an as-of-yet unmodeled impact of enhanced flame speed due to underlying differential diffusion effects; nevertheless, the scaling is consistent in all cases. 

\begin{figure} [ht!]
    \centering

   \includegraphics[width=\linewidth]{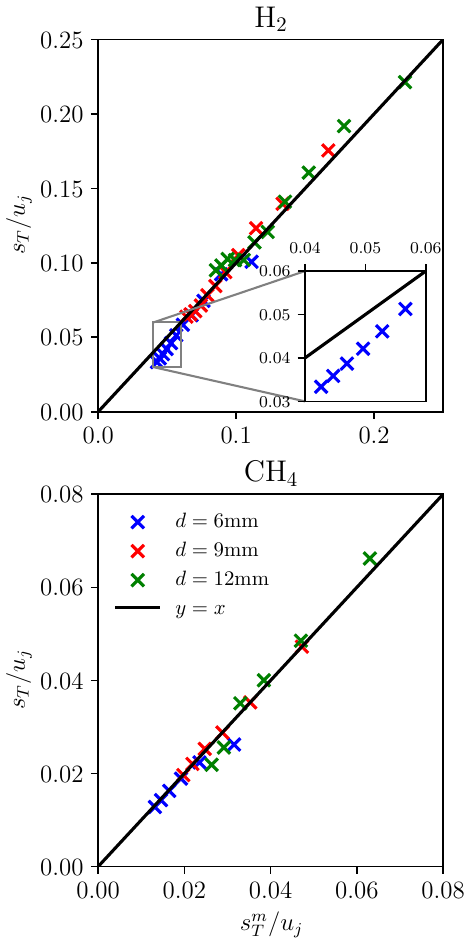}    
    \caption{Comparison between measured $s_{T} / u_{j}$ and modeled $s_{T}^{m} / u_{j}$ for \ce{H2} (top) and \ce{CH4} (bottom) flames. The solid line indicates the one-to-one correlation ($y = x$). The zoomed in panel for the \ce{H2} cases illustrates the relatively large difference ($\sim 20\%$) between the model and measured $s_{T}$ in the high Karlovitz cases.}
    \label{fig:st_model}
\end{figure}

This gives an overall correlation for the modeled normalized burning velocity, $\frac{s_{T}^{m}}{u_{j}}$,  as
\begin{equation}\label{eq:sT_model}
    \frac{s_{T}^{m}}{u_{j}} = \frac{s_{L}^{*}}{u_{j}} + \alpha_i Re_{F}^{\frac{1}{2}}Re_{j}^{-\frac{1}{2}}\frac{d_{j}}{l_{F}^{*}}.
\end{equation}
Figure~\ref{fig:st_model} illustrates this model by comparing the model turbulent burning velocity to the measured values, where a good agreement is observed in most cases. Minor discrepancies appear for \ce{H2} for $Ka_{\mathrm{eff}}^{*}>160$, with around a 20\% overestimation of the turbulent flame speed at these conditions. This potentially reflects a suppression of thermodiffusive effects not fully captured by the model. Overall, Fig.~\ref{fig:st_model} shows that the model reproduces the experimental scaling behavior with good consistency across a wide range of turbulent conditions ($Re_{j}, Ka_{eff}^{*}$) for both fuels.

\subsection{Flame shape and length\label{subsec:hf_s}}

\begin{figure} [ht!]
    \centering

    \includegraphics[width=0.85\linewidth]{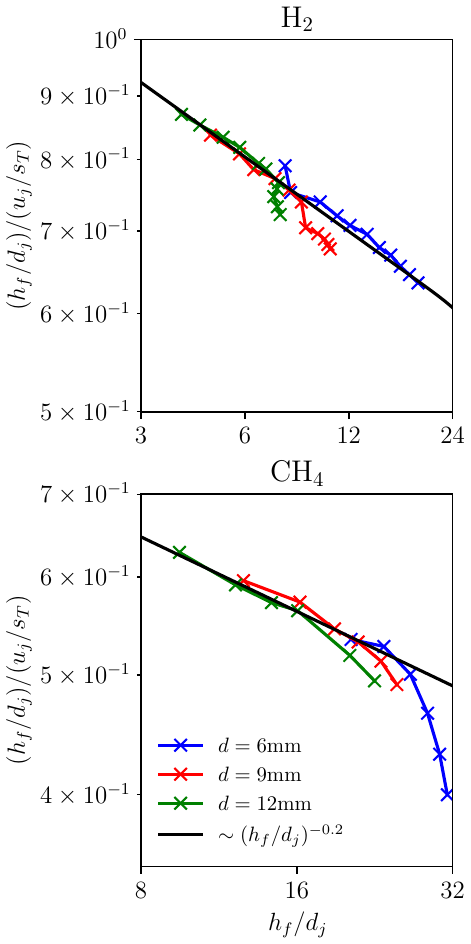}
    \caption{Scaling of the normalized shape factor with $h_{f}/d_{j}$ for \ce{H2} (top) and \ce{CH4} (bottom).}
    \label{fig:shape_scaling}
\end{figure}

Since the total burning rate directly governs the visible flame extent, it is expected that the normalized flame length, $h_{f}/d_{j}$, should scale inversely with $s_{T}/u_{j}$ according to Eq.~(\ref{eq:st_hf}). Therefore, the following section investigates the dependence of the measured flame length, $h_{f}$, and flame shape on $Re_{j}$ and fuel type, and assesses whether a single fuel-dependent constant can consistently describe both the mean geometric aspect of the turbulent flame structure. The mean flame shape is quantified through the ratio
\begin{equation}
\beta_i = \frac{h_f/d_j}{u_j/s_T},
\end{equation}
which is shown in Fig.~\ref{fig:shape_scaling} as a function of the normalized flame length. As $A_0$ changes, the average flame structure adjusts, leading to systematic changes in the observed flame geometry, from conical to columnar structures. For both fuels, $\beta_i$ follows a consistent power-law decay, with different fuel-dependent constants. This behavior can be described by
\begin{equation}\label{eq:shape_scaling}
\beta_i = \gamma_i \left(\frac{h_f}{d_j}\right)^{-0.2},
\end{equation}
where $\gamma_i$, the shape factor, is 0.98 for \ce{CH4} and 1.15 for \ce{H2}. Both \ce{H2} and \ce{CH4} follow a consistent power–law behavior with an effective slope of -0.2, indicating that the global flame geometry scales similarly across the investigated range. As the flame length increases, the flame shape evolves, with the effective power-law exponent $p$ decreasing with increasing normalized flame length. For \ce{H2}, the data from the three nozzle diameters collapses well along the $(h/d_{j})^{-0.2}$ trend. For $\ce{CH4}$, the data collapse well for flames shorter than 20 $d_{j}$, but deviations appear at larger flame lengths. Beyond this threshold, the flame rapidly approaches a cylindrical shape, more quickly than predicted by the power law before this point.

\begin{figure}[ht!]
    \centering

    \includegraphics[width=0.85\linewidth]{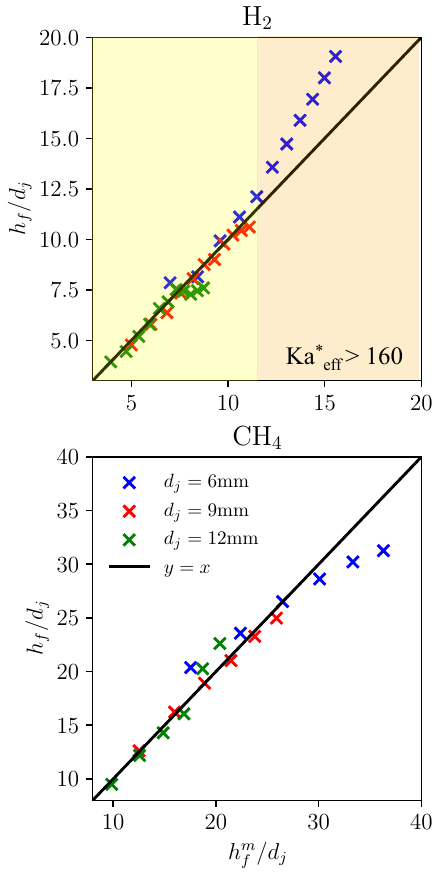}
    \caption{Comparison between modeled (${h^m_{f}}/{d_{j}}$) and measured ($h_{f}/d_{j}$) flame lengths for \ce{H2} (top) and \ce{CH4} (bottom).}
    \label{fig:h_model}
\end{figure}

From this scaling, an overall correlation for the normalized flame length model, $h^{m}_{f}/d_{j}$, can be derived as
\begin{equation}\label{eq:l_model}
    \frac{h^{m}_{f}}{d_{j}} = \left[\gamma_{i}^{-1}\left(\frac{s_{L}^{*}}{u_{j}} + \alpha_{i}Re_{F}^{\frac{1}{2}}Re_{j}^{-\frac{1}{2}}\frac{d_{j}}{l_{F}^{*}}\right)\right]^{-\frac{5}{6}},
\end{equation}
and the corresponding comparison with experimental data is shown in Fig.~\ref{fig:h_model}, where an overall good agreement is observed between the model and the measured flame height for the shorter flames, while deviations increase for longer ones. For \ce{H2}, the deviation in the $d_{j} = \qty{6}{\milli\meter}$ case is consistent with the rise of Karlovitz number ($Ka_{\rm{eff}}^{*} > 160$), where the model slightly overestimates the turbulent burning velocity, and hence underestimates the flame length. Conversely, larger nozzles ($d_{j} = \qty{9, 12}{\milli\meter}$) follow the scaling model closely, confirming its validity for more moderate Karlovitz numbers. For \ce{CH4}, the deviation from the model primarily arises from geometric effects: as $Re_{j}$ increases, the flame transitions more rapidly toward a cylindrical shape than predicted by the model when $h_{f}/d_{j} > 20$. This leads to an underestimation of the flame height in the small nozzle case ($d_{j} = \qty{6}{\milli\meter}$). 

In conclusion, the proposed scaling framework captures the turbulent burning velocity and flame length across fuels with different $Le$, with only $\alpha_i$ varying between \ce{H2} and \ce{CH4}. It confirms that classical scaling laws can be extended to lean \ce{H2} flames through appropriate adjustment of scaling coefficients. While some recent studies incorporate explicit Karlovitz-number-dependent corrections to account for the coupled influence of turbulence and thermodiffusive effects in \ce{H2} flames~\cite{howarth2023thermodiffusively, chu2025extended}, the consistent evaluation of Karlovitz number remains challenging in experimental datasets. In this context, the present framework provides a systematic way to represent fuel diffusivity effects in turbulence–flame interactions and to calibrate or validate turbulent combustion models using physically based, fuel-specific parameters rather than arbitrary empirical values.

\section{Conclusions and future perspectives\label{sec:end}} 

A comprehensive and consistent experimental dataset was obtained for premixed \ce{H2}/air and \ce{CH4}/air jet flames, enabling a comparison of turbulent burning velocity and flame length at fixed equivalence ratios over a wide range of Reynolds numbers and jet diameters. Damk\"ohler's small-scale limit turbulent burning velocity was found to model both sets of flames well (Eq.~\eqref{eq:jet_st_model}). Although the scaling with respect to the Reynolds number and diameter of both flames was the same, the flame speed factor $\alpha$ for both flames was notably different, with \ce{H2} ($\alpha_{\ce{H2}} = 0.28$) exhibiting a much larger value than \ce{CH4} ($\alpha_{\ce{CH4}} = 0.036$).  
This behavior may reflect interactions between turbulence and thermodiffusive instabilities, which affect the model constants. The model slightly overestimates the turbulent burning velocity at the highest Karlovitz numbers in the \ce{H2} cases (high $u_{j}$, small $d_{j}$), which may reflect a suppression of thermodiffusive effects in this regime. 
The mean flame shape changes with increasing flame length, exhibiting a gradual transition from an approximately conical to a cylindrical structure. This transition is captured using an empirical model (Eq.~\eqref{eq:shape_scaling}). The corresponding scaling coefficient differs between the two fuels, although the difference is relatively small ($\gamma_{\ce{H2}} = 1.15$, $\gamma_{\ce{CH4}} = 0.98$), indicating that \ce{H2} flames are slightly more compact, even though both fuels follow an identical power-law transition. By combining the shape-factor model with the turbulent burning velocity, a unified model for the flame length is obtained (Eq.~\eqref{eq:l_model}).
 This model shows good agreement across both flames in a wide range of conditions. Deviations from this model can be associated with the overestimation of the turbulent burning velocity for the high Karlovitz number $\ce{H2}$ cases, and the more rapid decay to a cylindrical shape for the very long $\ce{CH4}$ cases than the empirical shape model predicts. Future work will investigate the robustness of the parameters $\alpha$ and $\gamma$ across a broader range of equivalence ratios and operating conditions. The present framework will also be extended to \ce{CH4}/\ce{H2} blends in order to assess its applicability to mixed-fuel systems. Furthermore, the relationship between these constants and existing models for turbulent burning velocities of thermodiffusively unstable premixed flames will be systematically examined~\cite{howarth2022empirical,howarth2023thermodiffusively,hunt2025thermodiffusively}.

\acknowledgement{CRediT authorship contribution statement} \addvspace{10pt}

{\bf AM}: Writing – original draft, Methodology, Performed experiments, Formal analysis, Data curation, Conceptualization; {\bf TLH}: Writing – original draft, Methodology, Formal analysis, Data curation, Conceptualization; {\bf MC}: Writing – review \& editing, Methodology, Performed experiments, Formal analysis, Data curation, Conceptualization; {\bf FC}: Writing - review \& editing, Data curation; {\bf JB}: Writing - review \& editing, Supervision, Methodology, Conceptualization; {\bf MG}: Writing -review \& editing, Supervision, Methodology, Conceptualization; {\bf HP}: Writing - review \& editing, Supervision, Funding acquisition, Conceptualization.

\acknowledgement{Declaration of competing interest} \addvspace{10pt}

The authors declare that they have no known competing financial interests or personal relationships that could have appeared to influence the work reported in this paper.

\acknowledgement{Acknowledgments} \addvspace{10pt}

The ENCODING project has received funding from the European Union’s Horizon Europe research and innovation programme under the Marie Skłodowska-Curie grant agreement No 101072779. The results of this publication/presentation reflect only the author(s) view and do not necessarily reflect those of the European Union. The European Union cannot be held responsible for them. TLH acknowledges financial support from Deutsche Forschungsgemeinschaft (DFG) within the project (ID: 516338899) IRTG 2983 Hy-Potential. FC and JB acknowledge financial support form DFG within the project (ID: 523880888) SPP 2419. MC, MG, and HP acknowledge financial support from the European Research Council (ERC) Advanced Grant (HYDROGENATE, ID: 101054894).

\footnotesize
\baselineskip 9pt

\clearpage
\thispagestyle{empty}
\bibliographystyle{proci}
\bibliography{PROCI_LaTeX}


\newpage

\small
\baselineskip 10pt


\end{document}